# Size effect on polarization bremsstrahlung induced by electrons scattered on xenon clusters

Yu.S. Doronin, A.A. Tkachenko, V.L. Vakula, G.V. Kamarchuk

*B. Verkin Institute for Low Temperature Physics and Engineering*
*of the National Academy of Sciences of Ukraine*
*47 Nauky Ave., Kharkiv, 61103, Ukraine*

**Abstract**

We measured spectral distribution of absolute differential cross section of both ordinary and polarization bremsstrahlung for 0.3–1.0 keV electrons scattered on atoms and substrate-free nanoclusters of xenon. Clusters were produced in a supersonic gas jet expanding adiabatically into a vacuum. An original method based on absolute measurements of intensity of the atom and cluster emission in the VUV and USX spectral regions was used to determine the cluster density in the scattering area. The bremsstrahlung arising from scattering on clusters had a polarization component which dominated the differential cross section. For the first time cluster size effect on the formation of the polarization bremsstrahlung was found for xenon.

## Introduction

During the last decades, a lot of attention has been paid to the experimental and theoretical research of continuous bremsstrahlung (BS), which resulted in the development of several theoretical methods of analysis, among which the theory based on Refs. [1-2] occupies a prominent place. Methods and results of calculations for bremsstrahlung which, in addition to ordinary bremsstrahlung (OBS), has a polarization component (PBS), were published in Refs. [3-5]. Recently, on the basis of these methods, the BREMS code was developed to create a library of spectra and functions of the BS shape for many chemical elements and a wide range of incident electron energies [6].

Experimentally obtained BS cross sections for different energies of electrons scattered on thin C, Al, Te, Ta, and Au targets [7] revealed that, as the energy of BS photons decreases, a significant discrepancy arises between the theoretical and experimental results. This means that absolute differential BS cross sections for rare gases are of great interest, since they display substantial difference between the measured and calculated differenetial BS cross sections thanks to the pronounced PBS component observed in the BS spectra of these model substances [8-11].

The aim of the present work was to study in detail the process of formation of OBS and PBS induced by scattering of intermediate-energy electrons on xenon clusters and find the absolute differential BS cross sections for different cluster sizes. Obtaining such data may be very important for the development of the theory of bremsstrahlung covering both atoms and clusters.

## Experiment

The experimental method used in this work was described in Ref. [10]. The experimental setup consisted of an X-ray tube with a supersonic gas jet used as an anode and an X-ray spectrometer. A supersonic conical nozzle formed a jet that flowed adiabatically into a vacuum chamber. A beam of clusters and/or atoms, depending on the conditions of the experiment, was produced at the nozzle outlet. We varied the cluster size from 400 to 12000 atoms per cluster by varying the gas temperature (from 180 to 500 K) and pressure (from 0.03 to 0.1 MPa) at the nozzle inlet. The average cluster size was determined experimentally by the electron diffraction method [12, 13]. An electron beam crossed the xenon jet at a distance of 10 mm from the nozzle outlet. The electron energy was 0.3–1.0 keV, while the electron current was kept constant at 20 mA. The bremsstrahlung arising upon excitation of electrons was transformed into a spectrum by a RSM-500 monochromator spectrometer and recorded by a proportional counter.

The spectral distribution of the flux density of the bremsstrahlung generated in the photon energy range of 70–200 eV was determined in absolute units in accordance with a method [10] developed earlier. A supersonic argon jet was used as a calibrated source of radiation.

**Results**

Figure 1 shows BS spectra recorded in absolute units for 0.7-keV electrons scattered on xenon clusters with 400 and 12000 atoms per cluster. Similar behaviour of the BS spectral intensity was observed for all of the studied sizes of xenon clusters.

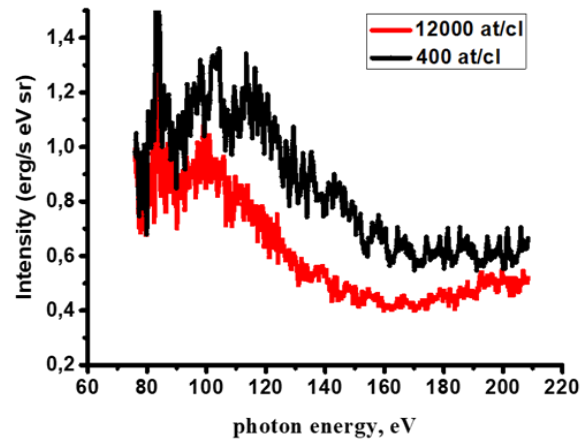

Fig. 1. Spectral intensity distribution of total BS in absolute units for 0.7-keV electrons scattered on xenon clusters with 400 and 12000 atoms per cluster.

Double differential cross section of total BS $\frac{d^2\sigma^{BS}}{d\omega d\Omega}$ (barn/eV·sr) was calculated using the following relation:

$$\frac{d^2\sigma^{BS}}{d\omega d\Omega} = \frac{I^{BS}(\hbar\omega)}{n_{cl}n_e v \hbar\omega} \cdot \frac{1}{A}.$$

Here $I^{BS}(\hbar\omega)$ (erg/s eV) is the BS intensity for photon energy $\hbar\omega$ (eV) in the solid angle $d\Omega$; $n_{cl}$, $n_e$ (cm$^{-3}$) are the densities of clusters and incident electrons, respectively; $v$ (cm/s) is the electron velocity; $A$ (cm$^3$) is the volume of the excitation area.

The size and number of xenon clusters interacting with electrons were measured using the original approach of Ref. [14].

Figure 2 shows spectral dependencies of differential BS cross sections for xenon atomic and cluster targets measured in the photon energy range 70–200 eV for 0.7-keV electrons. These data were obtained for clusters with 400 and 12000 atoms per cluster.

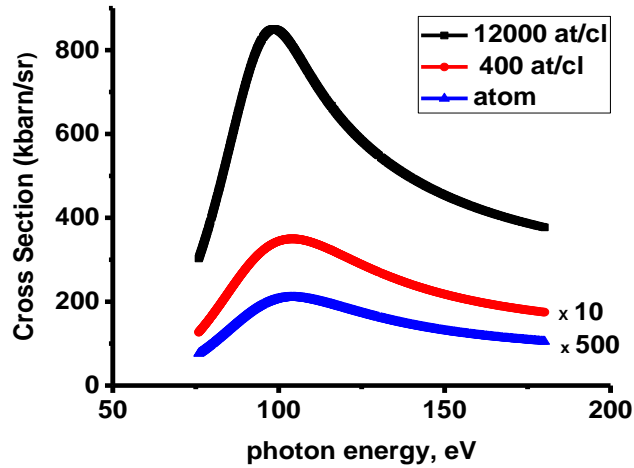

Fig. 2. Absolute differential BS cross section for 0.7-keV electrons scattered on Xe atoms and Xe clusters of different sizes.

The spectrum exhibits a well-defined broad maximum in the photon energy range of 80–155 eV, which is associated with the polarization bremsstrahlung. The BS cross section shape virtually does not change while its numerical value increases significantly with increasing cluster size. Analysis of the data in Fig. 2 makes it possible to estimate the contribution of the ordinary bremsstrahlung to the cross section of total bremsstrahlung for xenon clusters of different size. The obtained absolute double differential BS cross sections for total bremstrahlung (OBS+PBS), ordinary bremsstrahlung (OBS), and polarization bremsstrahlung (PBS), as well as some quantitative characteristics for xenon clusters in a wide range of their size are given in Table 1.

Table 1. Absolute double differential cross sections of OBS and PBS for xenon clusters.

| $T_0$, K | n, at/cl | Cluster density 1/cm$^3$ | Cluster radius, $R$(Å) | $R^2$, (Å$^2$) | Condensate fraction, % | Cross section (кbarn/sr) PBS+OBS | Cross section (кbarn/sr) OBS |
|---|---|---|---|---|---|---|---|
| 180 | 12000 | 2.8·10$^{11}$ | 50 | 2500 | 66 | 830 | 400 |
| 190 | 10000 | 3.1·10$^{11}$ | 47 | 2209 | 61 | 711 | 376 |
| 200 | 8000 | 3.8·10$^{11}$ | 44 | 1936 | 57 | 600 | 350 |
| 215 | 6000 | 4.3·10$^{11}$ | 39 | 1521 | 51 | 466 | 271 |
| 235 | 4200 | 5.1·10$^{11}$ | 35 | 1225 | 36 | 346 | 204 |
| 260 | 2800 | 5.2·10$^{11}$ | 31 | 961 | 29 | 260 | 152 |
| 320 | 1200 | 6.7·10$^{11}$ | 23 | 529 | 16 | 124 | 75 |
| 350 | 800 | 5.5·10$^{11}$ | 20.3 | 412 | 8.8 | 84 | 50 |
| 400 | 400 | 3.2·10$^{11}$ | 16.0 | 256 | 3.5 | 32 | 20 |
| 500 | atoms | - | 2.0 | 4 | - | 0.4 | 0.25 |

Figure 3 shows dependencies of absolute differential cross section of total bremsstrahlung (OBS+PBS) and ordinary bremsstrahlung (OBS) on square cluster radius $R^2$, which is proportional to the cluster geometrical cross section.

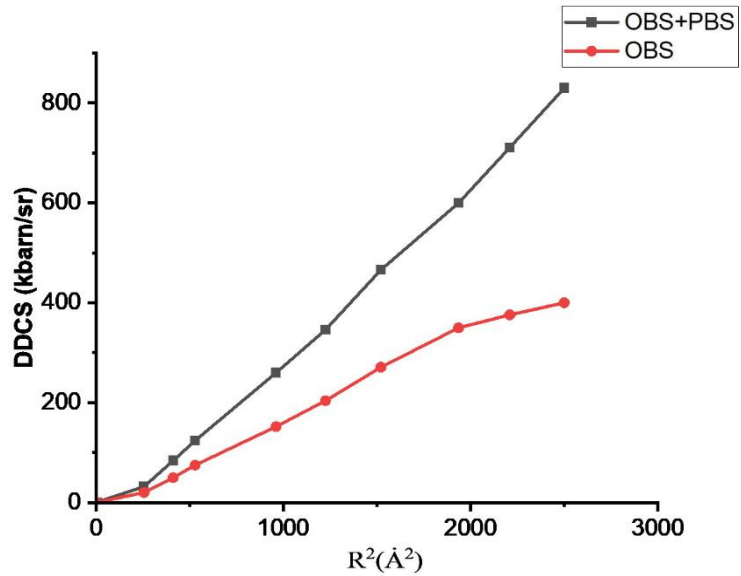

Fig. 3. Dependencies of the absolute double differential cross-section of total bremsstrahlung (OBS+PBS) and ordinary bremsstrahlung (OBS) on the square radius of xenon clusters.

It can be seen from Fig. 3 that differential cross sections, both OBS+PBS and OBS, increase with increasing cluster size. For clusters with 3000 and more atoms per cluster (square radii of 970 and more $Å^2$), the contribution of the polarization component to the differential cross section is the dominating one. This may be related to the many-particle interactions and interference of the atom contributions to the formation of the cluster bremsstrahlung. The obtained results show that in the case of clusters the cooperative phenomena have a significant effect on the main characteristics of bremsstrahlung.

**Conclusions**

Absolute double differential bremsstrahlung cross sections were obtained for electrons scattered on xenon clusters and a size effect on the polarization bremsstrahlung cross section was observed for the first time. These results are important for the development of bremsstrahlung theory and can be used in plasma diagnostics, astrophysics, and nanocluster physics.